\documentclass[10pt,a4paper]{article}
\usepackage[left=20mm, top=20mm, right=20mm, bottom=20mm, nohead, foot=10mm]{geometry}
\usepackage[utf8]{inputenc}
\usepackage[english]{babel}

\usepackage[affil-it]{authblk}
\usepackage{multicol}
\usepackage{caption}

\usepackage[hidelinks]{hyperref}
\usepackage{booktabs}
\usepackage{enumitem}
\usepackage{amsmath}
\usepackage{amssymb}
\usepackage{amsfonts}
\usepackage{mathrsfs}
\usepackage{multicol}
\usepackage{graphicx}
\usepackage{float}
\usepackage{xcolor}
\usepackage{empheq}

\usepackage[style=numeric,sorting=none,giveninits=true,isbn=false,alldates=year,doi=false,eprint=false,sortcites]{biblatex}
\renewbibmacro{in:}{}
\AtEveryBibitem{\clearfield{number}\clearfield{issue}}

\usepackage{titlesec}
\titleformat{\section}[block]{\normalfont\sffamily}{}{.5em}{\bfseries}
\titleformat{\subsection}[block]{\normalfont\sffamily}{}{.5em}{\bfseries}

\newenvironment{FigureOneColumn}
  {\par\medskip\noindent\minipage{\linewidth}\captionsetup{type=figure}}
  {\endminipage\par\medskip}

\begin{document}

\title{\sffamily Non-Abelian Vortices in Magnets}

\author[1]{\normalsize Filipp N. Rybakov\thanks{philipp.rybakov@physics.uu.se}}
\author[1,2]{\normalsize Olle Eriksson}

\affil[1]{\small
Department of Physics and Astronomy, Uppsala University, Box-516, Uppsala SE-751 20 Sweden
}
\affil[2]{\small
School of Science and Technology, \"Orebro University, SE-701 82 \"Orebro, Sweden
}

\date{}

\clubpenalty=10000
\widowpenalty=10000

\maketitle
\vspace{-3.0\baselineskip}

\renewcommand{\abstractname}{}
\begin{abstract}  
We show that non-Abelian vortices can exist in magnetic materials. These are singularity-free textures, are described by spin-lattice and field theory models, and we demonstrate that typical magnetic materials can be quite suitable for their realization and observation. We give a topological classification of these vortices and reveal their connection with Abelian topological structures, such as usual vortices, merons and skyrmions. We also describe how non-Abelian magnetic vortices can carry topologically protected information and present reasons for the advantage of such information carriers, as compared to Abelian entities.
\end{abstract}

\begin{multicols}{2}[]

\section*{Introduction}

In his famous review in 1979~\cite{RevModPhys.51.591}, Mermin characterized the non-Abelian (non-commutative) topological states in ordered media as ``the most interesting feature yet to emerge from the topological approach''. At that time, expectations were directed towards biaxial nematics~\cite{PoenaruToulouse1977}. Nowadays, the interest in non-Abelian systems is much broader. For example, there are non-Abelian anyons for topological quantum computation~\cite{RevModPhys.80.1083}, vortex strings with non-Abelian internal zero modes for high energy physics~\cite{Hanany_2003, AUZZI2003187}, and vortices in cyclic phase of spinor Bose-Einstein condensate~\cite{PhysRevLett.103.115301}.

In this work, we show that non-Abelian vortices also can exist in magnetic materials. Moreover, these vortices can be characterized as the most strongly non-Abelian, since they are described by the free group as having the fewest relations between their generators~\cite{ClimenhagaKatok}. 
We analyze the potential of non-Abelian magnetic vortices for information technology, and highlight their advantage, since they provide topological protection for all information, rather than individual bits, as in Abelian cases. Relevant materials parameters as well as geometrical considerations are also presented.

\section*{Model and topological analysis}

\begin{figure*}[t]
	\centering
	\includegraphics[width=14.4cm]{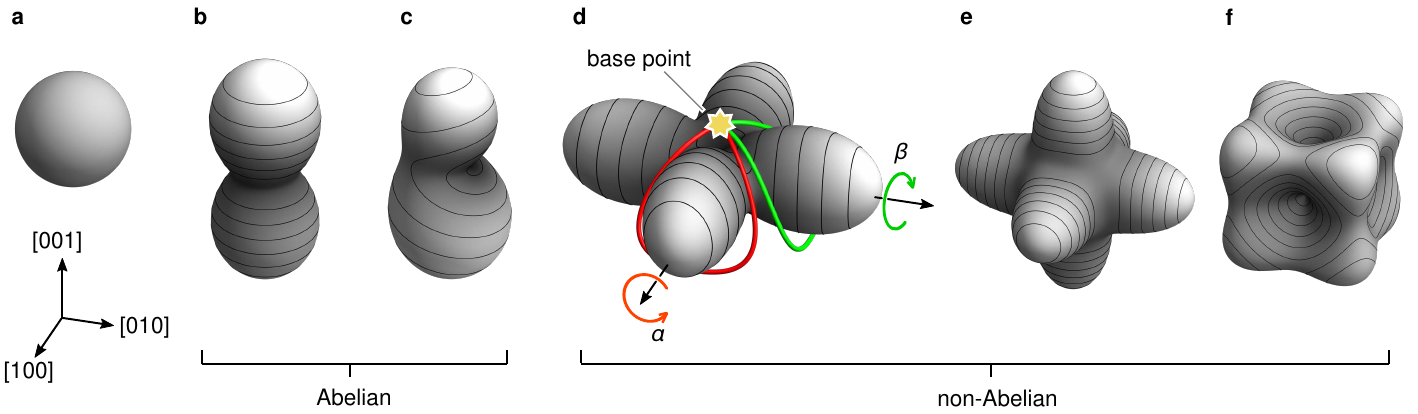}
	\caption{\small
Energy landscapes (surfaces) and contours of equal energy (lines) for potential term. 
The surfaces are described by parametric equation ${\mathbf{r} = \mathbf{s} f(\mathbf{s})}$, where parameterizing vector $\mathbf{s}$ takes on all possible values on the spin sphere, and the functions $f$ are described by Eq.~\ref{aniso}.
\textbf{a}, trivial isotropic case corresponding to zero values of all parameters in Eq.~\ref{aniso}. 
\textbf{b}, ${\mathrm{k}_\text{u} = 0.5}$. 
\textbf{c}, ${\mathrm{k}_\text{u} = 0.5}$ and ${\mathbf{h} = (0, 0.2, 0.1)}$. 
\textbf{d}, ${\mathrm{k}_\text{t1} = \mathrm{k}_\text{t2} = \mathrm{k}_\text{t3} = 0.1}$. 
\textbf{e}, ${\mathrm{k}_\text{c1} = -0.4}$ and ${\mathrm{k}_\text{c2} = 0.9}$.
\textbf{f}, ${\mathrm{k}_\text{c1} = 0.5}$.
}
	\label{fig1}
\end{figure*}

Consider the basic Hamiltonian~\cite{Aharoni_book} of either the lattice model of classical spins (Eq.~\ref{H_a}) or the corresponding continuum field model (Eq.~\ref{H_b}):   
\begin{subequations}
\begin{align}
H_\text{a} &=  
- J \sum_{\left\langle ij\right\rangle } \mathbf{s}_i  \cdot  \mathbf{s}_j 
 + 
K_\text{a}\!\sum_{i}f(\mathbf{s}_i), \label{H_a}
\\
H_\text{b} &= \int d{\mathbf{r}}\ 
\Bigg(\mathcal{A}\,\sum_{i=x,y,z}|\nabla m_i|^2 + K_\text{b} f(\mathbf{m})\Bigg),  
\label{H_b}
\end{align}
\label{H}
\end{subequations}

\noindent
where $\left\langle ij\right\rangle$ denote summation over all nearest-neighbour pairs, here chosen to be for a simple cubic lattice with lattice constant~$a$. 
This type of lattice was chosen in the calculations as the simplest one for a theoretical consideration. However, the generalization to various two-dimensional (2D) and three-dimensional (3D) lattices is straightforward. 
The parameter $J$ is the exchange coupling constant and $K_\text{a}$ is the common prefactor describing the energy reflecting the magnetic anisotropy. Similarly, for the continuum model, $A$ is the exchange stiffness and $K_\text{b}$ the common prefactor for the anisotropy energy density. 
For convenience, we consider the case when the two models of Eq.\ref{H} give similar excited state energies in the long-wavelength limit, so that ${A = J/(2a)}$ and ${K_\text{b} = K_\text{a}/a^3}$. 
In both cases, the order parameter space is the unit sphere $\mathbb{S}^2$, which we will also refer to as the spin sphere; so that for Eq.~\ref{H_a} we have $|\mathbf{s}_i|=1$ and for Eq.~\ref{H_b} we have $|\mathbf{m}(\mathbf{r})|=1$.  

The key ingredient of our analysis is the function $f$ -- the potential term, which defines the anisotropy profile of the spin direction:
\begin{subequations}
\begin{equation}
f(\mathbf{s}) = \text{const} -\  \mathbf{h}\cdot\mathbf{s}\  +  \nonumber\\
\end{equation}
\begin{empheq}[left = \empheqlbrace]{align}
&\mathrm{k}_\text{u} s_z^2,  \label{a_u} \\ 
&\mathrm{k}_\text{t1}(s_x^2 + s_y^2) + \mathrm{k}_\text{t2}(s_x^2 + s_y^2)^2 + \nonumber \\ 
&\qquad\qquad\qquad\quad\enspace\mathrm{k}_\text{t3}(s_x^4 - 6 s_x^2 s_y^2 + s_y^4), \label{a_t} \\ 
&\mathrm{k}_\text{c1}(s_x^2 s_y^2 + s_y^2 s_z^2 + s_z^2 s_x^2) + \mathrm{k}_\text{c2}(s_x s_y s_z)^2, \label{a_c}
\end{empheq}
\label{aniso}
\end{subequations}
where $\mathbf{h}$ is an applied magnetic field (considered to be dimensionless in our analysis) and the parameters $k_\text{u}$, $k_\text{t}$ and $k_\text{c}$ correspond to the cases of uniaxial, tetragonal, and cubic crystals, respectively~\cite{Kirchmayr_1996, Hubert}. 
Other cases of symmetries besides those listed in Eq.~\ref{aniso} are also possible, as well as higher orders of expansions~\cite{Aharoni_book}.
For simplicity, and without loss of generality, we assume that the main crystallographic directions, $[100]$, $[010]$ and $[001]$ coincide with Cartesian coordinates $(x,y,z)$ defining the orientation of the magnetic body. 
The generalization to the case of arbitrary mutual orientation is straightforward.
If dipole-dipole interactions (DDI) are significant, the corresponding local leading term should be taken into account in Eq.~\ref{aniso}, due to the ``shape anisotropy'' effect~\cite{doi:10.1098/rspa.1997.0013, Hubert, doi:10.1098/rspa.2016.0666}. 
Figure~\ref{fig1} shows some typical examples of energy surfaces for different potential terms.
The minima of $f(\mathbf{s})$ are usually related to the research focus because they determine the ground state, i.e. easy axis orientation, of the system.
In contrast, our approach is based on the analysis of maximums that can be realized in the center of a vortex core.

First, it is important to note that for topologically protected vortices, generally speaking, it is not required that the space of admissible values of the order parameter be nontrivial, such as the spin sphere. 
Indeed, for example in the Ginzburg-Landau theory~\cite{GL1950}, the relevant space is the complex plane, $\mathbb{C}$. Moreover, a potential Landau term in the form of a Mexican hat has a maximum where the order paramater is zero. Consequently, the system tends to reduce the size of domains with zero-valued order parameter, ultimately to points in a 2D situation or lines in a 3D case. Surrounding such points with a closed loop, one obtain that the order parameter along this loop does not contain such points and the corresponding classifying map is $\mathbb{S}^1\rightarrow\mathbb{C}\backslash\{0\}$, where the target space is simply a punctured plane. 
Therefore, the topological fundamental group is $\pi_1(\mathbb{C}\backslash\{0\})$~=  $\pi_1(\mathbb{R}^1\times\mathbb{S}^1)$~=  $\pi_1(\mathbb{R}^1)\times\pi_1(\mathbb{S}^1)$~=  $\pi_1(\mathbb{S}^1)$~= $\mathbb{Z}$. 
That is, it is a group of integers that is Abelian.
The non-trivial fundamental homotopy group is proof that an isolated superconducting vortex is topologically protected and can be removed by pushing it to the sample boundary rather than any local perturbation. 

Similar arguments for deriving fundamental groups are applicable for magnets, where the space of the order parameter is nontrivial by itself. 
Consider the case of easy-plane anisotropy presented in Fig~\ref{fig1}(b). 
Alternatively, one can consider the case presented in Fig~\ref{fig1}(c), where an applied field somewhat breaks the $U(1)$ symmetry.
At this point, for our purposes, it is only important that there are two maxima. 
Let's denote the corresponding points on the spin sphere as $P_1$ and $P_2$ and calculate the fundamental group classifying vortices in the same manner as above (by puncturing the points corresponding to the most energetically expensive directions): $\pi_1(\mathbb{S}^2\backslash\{P_1, P_2\})$~= $\pi_1(\mathbb{S}^1)$ = $\mathbb{Z}$. 
This intermediate result is directly related to a number of remarkable phenomena in magnetism associated with Abelian vortices. Here are some examples. 
Elementary planar vortices, often studied in the limit ${\mathrm{k}_\text{u}\rightarrow\infty}$, have topological charges~${\pm 1}$. Thermal fluctuations can generate vortex-antivortex pairs, which are topologically trivial (${-1+1 = 0}$) and therefore can be compact. Stronger fluctuations can lead to a Berezinskii–Kosterlitz–Thouless phase transition associated with the dissociation of such a pair~\cite{Berezinskii_1970,Kosterlitz_1973,SvistunovBabaevProkofev}. 
Other examples are micromagnetic vortices~\cite{Landau_diamond,Feldtkeller1965,Behncke_incollection} and merons~\cite{PhysRevB.91.224407,PhysRevB.101.064408,Braun_incollection}. If the total topological charge of the vortices (merons) is not zero, then unboundedness in size is a consequence of the nontriviality of the corresponding element of the fundamental group~$\pi_1(\mathbb{S}^1)$, which in turn means that the magnetic texture must occupy the entire sample.

Next we consider cases where the number of maxima, $n$, is two or higher,  such as shown in Figs.~\ref{fig1}(d)-(f). For the map defining the fundamental group, the target space is an $n$-punctured sphere and it deformation retracts onto a bouquet of $n-1$ circles, see Ref.~\cite{Zubarev_YouTube} for an animation. The appropriate computation~\cite{ClimenhagaKatok} therefore gives 
\begin{align}
\pi_1(\mathbb{S}^2 \backslash & \{P_1, P_2, ..., P_n\} ) = \nonumber\\
&\pi_1(\underbrace{\mathbb{S}^1\vee\mathbb{S}^1\vee...\vee\mathbb{S}^1}_{n-1 \text{ circles}}) = F_S,\label{homot}
\end{align}
where $F_S$ is a free group of rank ${n-1}$, which is non-Abelian if ${n>2}$. 
This result also shows a connection with the above-mentioned usual magnetic vortices and merons corresponding to the special case of ${n=2}$ when the rank is equal to one and the free group is Abelian and isomorphic to integers, ${F_{\{1\}} = \mathbb{Z}}$.
Based on the analysis above, we can formulate a recipe for non-Abelian magnetic vortices, which consists in choosing a system in which the effective potential term has 3 or more local energy maxima. 

It should be noted that states that are topologically trivial under the fundamental group can, generally speaking, be skyrmions.  
A skyrmion can be formed by a vortex-antivortex pair in the non-Abelian case as well as in the Abelian one~\cite{Zhang2015, PhysRevLett.119.207201, PhysRevB.101.064408}.
Skyrmions formed from combinations of non-Abelian vortices and antivortices will nevertheless be Abelian, as long as their homotopy group is ${\pi_2(\mathbb{S}^2)=\mathbb{Z}}$.

\section*{Elementary vortices and multivortices}

\begin{figure*}
	\centering
	\includegraphics[width=14.4cm]{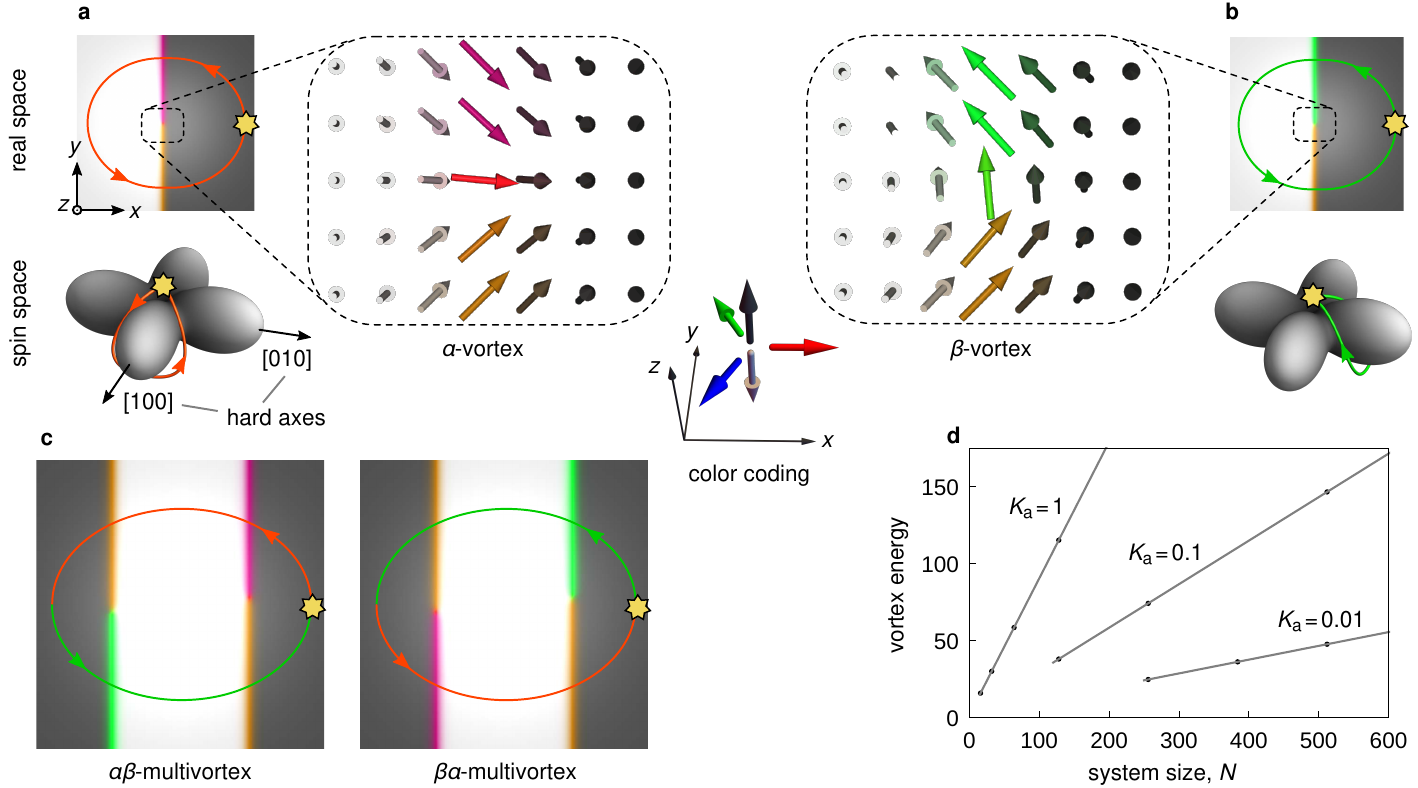}
	\caption{\small
Non-Abelian vortices in the model of a tetragonal magnet.
\textbf{a} and \textbf{b}, Structure of two elementary vortices corresponding to Fig.~\ref{fig1}(d).
\textbf{c}, Two topologically distinct variants of a multivortex consisting of $\alpha$-vortex and $\beta$-vortex.
\textbf{d}, Calculated dependence of the energy of an elementary vortex on the size of the system.
The star icon indicates the position of the base point. 
}
	\label{fig2}
\end{figure*}

Without loss of generality, but in order to demonstrate a concrete example, we consider the case in Fig.~\ref{fig1}(d). 
The presence of a single easy axis in this particular case makes the elements of non-Abelian vortices similar to typical domain walls and Bloch lines~\cite{MalozemoffSlonczewski_1979,Hubert}.
First, we choose a base (basic) point~\cite{HuHomotopyTheory} from the set of minima. 
Note that for the convenience of notation of homotopy groups, the base point may not be indicated explicitly, like in Eq.~\ref{homot}, but its presence is essential and implied~\cite{HuHomotopyTheory}. 
Let the base point be ${\mathbf{s}_0=(0,0,1)}$ in spin space and, at the same time, be a point located along the $\hat{\mathbf{e}}_x$ direction in real space (see the star icon on Fig.~\ref{fig2}(a)).
Four maxima mean that there are three generators of the group $F_S$, which we will denote by Greek letters, namely ${S=\{\alpha,\beta,\gamma\}}$, and the corresponding inverse is ${S^{-1}=\{\alpha^{-1}, \beta^{-1}, \gamma^{-1}\}}$.
Let $\alpha$ and $\beta$ correspond to one turn counterclockwise and clockwise around the peaks oriented towards $[100]$ and $[010]$ respectively, see Fig.~\ref{fig1}(d). 
For completeness of the description, one can choose for example one turn counterclockwise around the peak oriented towards $[\bar{1}00]$ as a $\gamma$.

Topological protection is guaranteed by Eq.~\ref{homot} and numerical calculations are not essential to prove this aspect. 
However, one has to carry out calculations in order to reveal more detailed information such as the profile and structure of the vortices, as well as their energies. 
The calculation of vortex solutions was carried out by the energy minimization method using Eq.~\ref{H_a}, for lattices of various sizes, ${N \times N}$, for ${J=1}$ and $K_\text{a}$ taking the values from $0.01$ to $1$. For the details on numerical technique see the Supplemental Material in Ref.~\cite{PhysRevLett.115.117201}. 
The structure of the static solutions found for elementary vortices is shown in Figs.~\ref{fig2}(a) and~(b).  
When rotating counterclockwise in the $xy$-plane around the core (red loop in Fig.~\ref{fig2}(a) and green loop in Figs.~\ref{fig2}(b)) the spins follow counterclockwise or clockwise on the spin sphere, making a turn around the $[100]$- or $[010]$-axis.
The texture is completely singularity-free, and the maximum angle between neighboring spins in the found solutions did not exceed $5^{\circ}$ for ${K_\text{a}=0.01}$ and $46^{\circ}$ for ${K_\text{a}=1}$.
Considering that Eq.~\ref{H_a} can be viewed as a finite-difference approximation of Eq.~\ref{H_b}, it can be concluded that the continuum model is applicable to the description of such vortices if the anisotropy energy is not extremely high.
The plot on Fig.~\ref{fig2}(d) shows the dependence of the energy of the vortex on the size of the system. 
This dependence is asymptotically linear, since at large distances from the core the structure is a collection of domain walls whose energy is proportional to their length.

Let us now turn to the case of many-vortex configurations to demonstrate noncommutative properties. By virtue of Eq.~\ref{homot}, different elements of the group $F_S$ correspond to different homotopy classes. Accordingly, since for the elements $\alpha\beta$ and $\beta\alpha$ it is true that $\alpha\beta\neq\beta\alpha$, there is no way for the corresponding multivortices to transform from one to another. 
Figure~\ref{fig2}(c) shows the numerical solutions obtained from Eq.~\ref{H_a}, on a 320$\times$320 lattice with ${K_\text{a}=0.1}$ for these two topologically distinct states, which, however are composed of the same type of elementary vortices. 
The cores of the vortices are repulsive, which is somewhat predictable based on purely topological arguments~\cite{PoenaruToulouse1977, RevModPhys.51.591}. 
When minimizing the energy, we used the initial guess by placing the vortex cores very close to each other.
The obtained solutions show that in equilibrium the cores are well separated, as is visible from the width of the central ``light'' domain in Fig.~\ref{fig2}(c).
However, the width of this domain can be reduced by applying a small field along the $z$-axis.
A further increase in the strength of the applied field lead to the scattering of cores in the opposite directions of the $y$-axis.

\section*{Topologically protected information}

Micromagnetic textures such as domain walls or skyrmions can serve as information carriers~\cite{doi:10.1126/science.1145799, Fert2013}.
In this regard, it is important to emphasize the potential advantages of non-Abelian vortices in such applications. 
Consider information storage capable of holding $N$ interacting carriers. 
From a topological point of view, their total charge is protected.
For the Abelian case, if each carrier can have a topological charge from a finite set of values, then the number of topologically distinct alternatives is simply proportional to $N$. 
Since the information measure is based on the logarithm of the number of alternatives~\cite{information_Hartley, information_Host}, the topologically protected information is proportional to $\log(N)$. 
However, in the case of non-Abelian vortices, due to non-commutativity, 
the number of topologically distinct alternatives is proportional to a number with $N$ in the exponent. Accordingly, the topologically protected information is proportional to $N$, see Fig.~\ref{fig3}(a). 

As a proof-of-principle, we performed 3D micromagnetic modeling (Eq.~\ref{H_b}) with material-specific constants to demonstrate how non-Abelian vortices could be utilized in information storage. 
We chose the constants close to those for Nd$_{2}$Fe$_{14}$B at temperatures about~200~K~\cite{doi:10.1063/1.339789, PASTUSHENKOV1997278}: $\mathcal{A}=10^{-11}$~J\,m$^{-1}$,
saturation magnetization ${M_\text{s}=1.6\cdot10^{6}}$~A\,m$^{-1}$, 
$K_\text{b}=10^{7}$~J\,m$^{-3}$, ${\mathrm{k}_\text{t1} = \mathrm{k}_\text{t2} = 0.375}$ and ${\mathrm{k}_\text{t3} = 0.15}$.
To reduce the demagnetizing fields, we consider a sample in the form of a rectangular lamella fabricated in such a way that the easy axis lies in the plane of the sample, see Fig.~\ref{fig3}(b).
The lamella size is $100\times50\times10$~nm.
Since the anisotropy energy reaches its maximum in the core of the vortex, it seems reasonable for our purposes to suggest a system where the vortices could be core-less. 
To realize this, we assume the presence of a narrow hole of size $80\times10$~nm along the lamella.
As a result, non-Abelian vortices are trapped, since their escape from the sample requires the energy-consuming nucleation of the core.
This confinement method is based on the same principle as for Abelian vortices, such as the capture of a superconducting vortex by a superconductor with a hole or the capture of a micromagnetic vortex in a magnetic ring~\cite{PhysRevLett.86.1102}.

The micromagnetic problem was solved by a finite-difference method described in Ref.~\cite{Zheng2021} using Excalibur software~\cite{excalibur}.
The mesh size was chosen to be 0.5~nm.
Figure~\ref{fig3}(b) shows one solution from the set of stable states corresponding to all possible values of the 4-bit memory register schematized in Fig.\ref{fig3}(c).
If, for example, we associate a binary ``0'' with $\alpha$ and ``1'' with $\beta$, and choosing as base-point a position on the right, short side of the device in Fig.~\ref{fig3}(b), a counterclockwise movement over the state in Figure~\ref{fig3}(b) would encode the content ``0100''. 
A longer pattern will hold proportionately more bits and again all information is topologically protected.

\begin{FigureOneColumn}
	\centering
	\includegraphics[width=7.2cm]{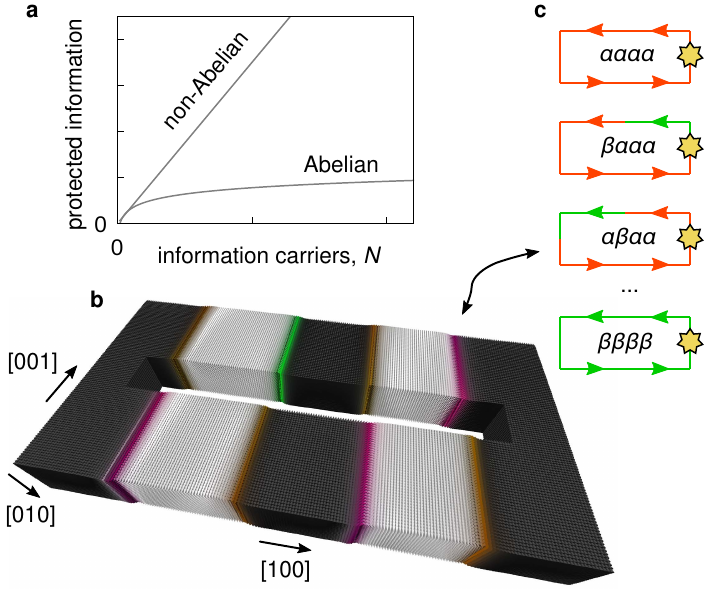}
	\caption{\small
Non-Abelian vortices for information processing.
\textbf{a}, Dependence of topologically protected information on the amount of carriers for non-Abelian and Abelian cases.
\textbf{b}, Structure of core-less $\alpha\beta\alpha\alpha$-multivortex in nanosample of tetragonal magnet.
\textbf{c}, Schematic representation of different multivortices associated with different states of 4-bit memory register. 
}
	\label{fig3}
\end{FigureOneColumn}

\section*{Discussion and Conclusion}

Being topologically protected, the vortices discussed here do not need any stabilizing interactions additional to Eq.~\ref{H}. 
However, accounting for additional interactions like DDI, competing exchange interactions (frustrations)~\cite{VILLAIN1959303, Garel_1980}, Dzyaloshinskii–Moriya interaction, RKKY interaction,  antichirality-related terms~\cite{PhysRevB.104.L020406, PhysRevLett.127.127204} and others opens up a number of additional possibilities. 
Thus, for example, additional interaction may be responsible for fixing core-core or core-edge distances.

For the experimental realization of non-Abelian magnetic vortices, various magnetic materials could be suitable, as long as the anisotropy is more sophisticated than uniaxial. 
Along with materials that have been studied for a long time, such as Fe$_{14}$Nd$_{2}$B~\cite{doi:10.1063/1.339789, PASTUSHENKOV1997278}, some 3$d$-5$d$ transition metal compounds may be suitable. A detailed selection of suitable material could be guided by data-mining efforts and data-bases that aim to identify functional magnetic materials ~\cite{PhysRevB.101.094407}.
To observe small numbers of non-Abelian magnetic vortices, the most common isotropic magnetic materials may be suitable. In this case, the required anisotropy may be induced by the shape of specially designed and fabricated microscopic samples~\cite{doi:10.1063/1.4948455, doi:10.1063/1.4942445}.

It seems interesting to pursue theoretical studies of non-Abelian magnetic vortices in other contexts than for information carriers. Especially their dynamics in three dimensions, which is governed by the Landau–Lifshitz–Gilbert equation. Collisions of non-Abelian vortex strings are highly non-trivial from general theories~\cite{RevModPhys.51.591, PoenaruToulouse1977, PhysRevLett.103.115301}, and simulations as well as experiments in magnetic materials could improve on the understanding of these systems in a general sense. 

In addition to data storage, we speculate that non-Abelian magnetic vortices can be considered for quantum computing. 

\section*{Acknowledgments}

The authors acknowledge support from the Swedish Research Council. O.E. also acknowledges support from the Knut and Alice Wallenberg (KAW) foundation, the ERC (FASTCORR project) and eSSENCE. 

\renewcommand*{\bibfont}{\small}
\printbibliography

\end{multicols}

\end{document}